\documentclass[11pt]{article}
\usepackage[margin=1in]{geometry}
\usepackage{amsmath,amssymb,amsthm}
\usepackage{graphicx}
\usepackage{booktabs}
\usepackage{bm}
\usepackage{hyperref}

\title{A Preliminary Theory of Infantile Dynamics}

\author{Lei Ma\thanks{Spiking Labs, \url{https://spikinglabs.com/}}}
\date{April 1, 2026}

\begin{document}
\maketitle

\begin{abstract}
We present a concise dynamical picture of infant-driven household chaos. The framework has three postulations: recurrent daily chaos, overall entropy growth in household organization, and transient local ordering episodes with switching rules (a volatile Maxwell-demon effect). We illustrate entropy growth with a two-region toy model (organizer vs. play area), where entropy production is nonnegative and long-time behavior is typically play-area dominated. We also model toy diffusion and curiosity-driven behavior, where novelty matters more than punishment in the short term, while gradual learning still occurs. 
\end{abstract}

\section{Introduction}
Household order in baby-occupied environments is typically temporary. To describe this repeatable cycle of cleanup and re-chaos, we use three postulations: recurrence of chaotic states, coarse-grained entropy growth, and transient time-varying local ordering episodes (a volatile ``baby demon'' effect). We instantiate these ideas with a two-region H-theorem toy model, an entropy-driven diffusion model, and a curiosity-dominant learning model.

The household state includes toy locations, food-particle distribution, caregiver stress, and infant acoustic output.

\section{The Macro-Trajectory of Household Entropy}
\subsection{Postulation 1: Infantile Recurrence Theorem (Poincar\'e Analogue)}

Household chaos is not a one-off event: the system repeatedly returns to states arbitrarily close to earlier ones. For example , even after temporary order is restored, familiar toy-scattering patterns and caregiver-state configurations reappear over time.

We state it in a measure-theoretic form analogous to the classical Poincar\'e recurrence theorem~\cite{Poincare1890,BARREIRA}. Let $(X,\Sigma,\mu)$ denote the household phase space, where $X$ is the set of admissible household microstates, $\Sigma$ is a $\sigma$-algebra of measurable state sets, and $\mu(X)<\infty$ is the finite accessible phase volume. Let $\Phi_t:X\to X$ be the baby--household flow map (continuous-time dynamics), and define a one-step map $T=\Phi_{\Delta t}$.

We assume (i) \emph{bounded dynamics}: trajectories remain in $X$ within time window large enough, $t \in [0, T_\infty)$, and (ii) \emph{conservative dynamics}: for every measurable $A\in\Sigma$, 
\begin{equation}
\mu\!\left(\Phi_t^{-1}(A)\right)=\mu(A),
\end{equation}
so phase-volume is preserved under time evolution.

For any measurable set $E\subseteq X$ with $\mu(E)>0$, recurrence means that for $\mu$-almost every $x\in E$, there exist infinitely many integers $n\ge 1$ such that
$T^n(x)$ means applying the map $T$ repeatedly $n$ times to the initial state $x$ (so $T^1(x)=T(x)$, $T^2(x)=T(T(x))$, etc.).
\begin{equation}
T^n(x)\in E.
\end{equation}
Equivalently, in metric form with distance $d$ on $X$, for $\mu$-almost every initial state $x$ and every tolerance $\epsilon>0$, there exists $n\in\mathbb{N}$ such that
\begin{equation}
d\!\left(T^n(x),x\right)<\epsilon.
\end{equation}

Here, $x$ is a full microstate (toy positions, food crumbs, caregiver stress, and acoustic mode), $n$ is the number of discrete observation steps of size $\Delta t$, and $\epsilon$ quantifies how close a return must be to count as recurrence. Thus $T^n(x)$ is the household state after $n$ update steps.

In practice, we model the dominant recurrence scale as
\begin{equation}
\tau_{\text{recurrence}}\approx 24\ \text{hours},
\end{equation}
where $\tau_{\text{recurrence}}$ is an empirical return-time statistic, not a universal constant and not identical for all initial states.

\subsection{Postulation 2: Macro-Entropy Growth Hypothesis}

We introduce Postulation 2 as a modeling hypothesis: under typical baby--household dynamics, the coarse-grained household entropy tends to increase over time unless sustained external work is applied (e.g., parental cleanup). To formalize this postulation, we use the Gibbs entropy over microstate probabilities~\cite{Gibbs1902,jaynes1965}:

\begin{equation}
S_{\text{house}}(t)=-k_B\sum_{i} p_i(t)\ln p_i(t),
\end{equation}
where $p_i(t)$ is the probability of microstate $i$ at time $t$. In this paper, a microstate specifies fine-grained toy locations (organizer bins, floor regions, under-furniture traps), food-particle placement, and caregiver-baby interaction mode. In the toy-and-baby context, entropy is low when most probability mass is concentrated on a few tidy configurations (e.g., toys mostly in organizers), and entropy increases as baby activity spreads probability across many mixed, cluttered configurations.

\subsubsection*{Example Under Postulation 2: Two-Region H-Theorem Model}
To connect this with irreversible mixing, we introduce an H-theorem analogue for a two-region room: region $O$ (toy organizer) and region $P$ (play area). Let $p_O(t)$ and $p_P(t)$ be the probabilities that a randomly chosen toy is in $O$ or $P$, with $p_O+p_P=1$.
Assume coarse-grained toy motion between the two regions is Markovian with transition rates $k_{OP}$ (organizer $\to$ play area) and $k_{PO}$ (play area $\to$ organizer):
\begin{equation}
\dot p_O=-k_{OP}p_O+k_{PO}p_P, \qquad \dot p_P=-k_{PO}p_P+k_{OP}p_O.
\end{equation}
Define the H-function directly as relative entropy to the stationary occupancy:
\begin{equation}
H(t)=p_O\ln\!\frac{p_O}{\pi_O}+p_P\ln\!\frac{p_P}{\pi_P},
\end{equation}
where $\pi_O=\frac{k_{PO}}{k_{OP}+k_{PO}}$ and $\pi_P=\frac{k_{OP}}{k_{OP}+k_{PO}}$. In the symmetric case $k_{OP}=k_{PO}$, this becomes $H=p_O\ln p_O+p_P\ln p_P+\ln 2$, i.e., the familiar two-state form up to an additive constant. As an alternative asymmetric case reflecting geometry, if the play area is much larger and toys are more likely to remain there, we expect $k_{OP}\gg k_{PO}$, hence $\pi_P\gg \pi_O$; the long-time state is then strongly play-area dominated with $p_P\approx \pi_P\gg p_O\approx \pi_O$.

Using the master equation and $\dot p_O+\dot p_P=0$, we obtain
\begin{equation}
\frac{dH}{dt}=\dot p_O\ln\!\frac{p_O\pi_P}{p_P\pi_O}
= -\left(k_{OP}p_O-k_{PO}p_P\right)\ln\!\frac{k_{OP}p_O}{k_{PO}p_P}\le 0,
\end{equation}
since $(a-b)\ln(a/b)\ge 0$ for $a,b>0$. Therefore, this same $H(t)$ never increases with time; equality holds only at stationarity $k_{OP}p_O=k_{PO}p_P$.

Defining entropy production through this dissipation,
\begin{equation}
\sigma(t)\equiv -k_B\frac{dH}{dt}\ge 0,
\end{equation}
we obtain the intended H-theorem statement for the two-region toy model.

With the H-theorem, we observe that once toys are released from the organizer into the play region, spontaneous dynamics overwhelmingly favor further mixing rather than perfectly reversing back into neat storage. In the asymmetric large-play-area case ($k_{OP}\gg k_{PO}$), this means the typical long-time household state is play-area dominated ($p_P\gg p_O$): most toys remain distributed across the play region, while only a small fraction resides in the organizer at any given time.

\subsubsection*{Entropy Driven Infantile Toy Diffusion}

As a direct consequence of the macro-entropy growth model, and an anternative view of the phenomenon, we discuss the toy diffusion process: toys initialized in a high-concentration storage region (e.g., one basket) diffuse via repeated infant--object collisions. The toy concentration field $C_{\text{toy}}(\mathbf{x},t)$ obeys~\cite{Fick1855}:
\begin{equation}
\frac{\partial C_{\text{toy}}}{\partial t}=D_{\text{baby}}\nabla^2 C_{\text{toy}},
\end{equation}
where $D_{\text{baby}}$ is the infantile diffusion coefficient in $\mathrm{m^2/s}$.

We additionally define a baby curiosity potential $\Phi_b(i)$ for object $i$ as its propensity to be relocated by infant interaction. Let $\Delta r_i=\lVert \mathbf{x}_i(t+\Delta t)-\mathbf{x}_i(t)\rVert$ denote displacement over interval $\Delta t$; then, in expectation,
\begin{equation}
\mathbb{E}[\Delta r_i] \propto \Phi_b(i),
\end{equation}
so larger curiosity potential implies farther toy transport under repeated encounters.

\begin{equation}
\mathbf{F}_{\mathrm{ent}}^{(i)} = T_{\mathrm{nursery}}\,\nabla S_{\mathrm{baby}}\!\left(\mathbf{x}_i\right),
\end{equation}
where $\mathbf{F}_{\mathrm{ent}}^{(i)}$ is the effective entropic force induced by baby activity, $T_{\mathrm{nursery}}$ is an ambient household temperature scale, and $S_{\mathrm{baby}}$ is the baby-generated entropy field. Objects with larger $\Phi_b(i)$ couple more strongly to this force and therefore exhibit larger expected displacement.

As a result, in open-plan apartments, diffusion approaches uniform floor occupancy by late afternoon, driven the effective baby entropic force.

\section{Statistical Fluctuations and Micro-Anomalies}
\subsection{Postulation 3: Volatile Maxwell Demon Anomaly}
Despite global entropy growth, infants can transiently lower \emph{local} entropy by acting as a time-varying Maxwell demon with changing selection rules. Let the active target class be $g(t)$: during one window the baby extracts yellow toys ($g=\text{yellow}$), and in a later window the baby switches to round toys ($g=\text{round}$). Writing $C_g$ for the concentration field of the currently targeted class, we model
\begin{equation}
\frac{dS_{\text{local}}}{dt}=
\begin{cases}
-D_y\,\nabla C_{\text{yellow}}, & 0\le t<\tau_1,\\[4pt]
-D_r\,\nabla C_{\text{round}}, & \tau_1\le t<\tau_2,\\[4pt]
R_{\text{reset}}, & t\ge \tau_2,
\end{cases}
\end{equation}
with $R_{\text{reset}}\gg 0$ denoting the high-entropy reset phase.

Order can appear briefly in multiple modes (first color-sorted, then shape-sorted), but rule-switching plus reset dynamics ultimately restore high local disorder.

\section{Asymptotic Moderation of the Three Postulations}

The three postulations describe the dominant short-horizon regime of baby--household dynamics. Over longer horizons, however, their practical strength is moderated by learning: recurrence remains, but the exact recurrent patterns evolve; entropy still tends to grow, but at a slower rate under routines; and local anomaly episodes become less disruptive. A simple mechanism for this transition is a gradual shift in the effective reward balance from novelty-dominant to mixed novelty--constraint control.

\subsection{Curiosity-Driven Learning Dynamics}

Infant exploration is primarily curiosity-driven rather than punishment-driven. In household settings, novelty and sensory richness dominate behavior selection, while punishment signals often have weak short-term control effect~\cite{SuttonBarto2018,OudeyerKaplan2007}.

A compact behavioral model is
\begin{equation}
\pi(a\mid s) \propto \exp\!\left(\beta\,[\alpha\,N(s,a)+R_{\text{social}}(s,a)-\lambda\,P(s,a)]\right),
\end{equation}
This indicates that the baby tends to choose actions that feel new or engaging, even if they were discouraged before. Here $N(s,a)$ is novelty value, $R_{\text{social}}(s,a)$ is interaction reward, and $P(s,a)$ is punishment cost; $\alpha$ is curiosity strength, $\lambda$ is punishment sensitivity, and $\beta$ controls action selectivity. In early stages, the regime $\alpha\gg\lambda$ explains why correction alone is often ineffective.

As development proceeds, repeated boundaries and predictable routines effectively increase the influence of constraint terms (larger effective $\lambda$ and structured $R_{\text{social}}$). Consequently, high-disruption actions become less frequent, and the postulations should be interpreted as early- to mid-stage asymptotic tendencies rather than immutable laws.

\section*{Acknowledgments}

This April Fools' paper was inspired by Yoyo. We thank Gemini 3.1 Pro for insightful discussions.

Lastly, we dedicate this work to all caregivers, who tirelessly act as the stabilizing forces in a highly volatile phase space, proving daily that love is the only force capable of overcoming the second postulation.

\bibliographystyle{unsrt}
\bibliography{references}

\end{document}